\documentclass[preprint,aps]{revtex4}
\usepackage{epsfig}
\begin{document}
\title{ON MATRIX SUPERPOTENTIAL AND THREE-COMPONENT NORMAL MODES}
\author{R. de Lima Rodrigues$^{(a)},$
A. F. de Lima$^{(b)},$\\
 E. R. Bezerra de Mello$^{(c)},$
and V. B. Bezerra$^{(c)}$\\
{}$^{(a)}$Centro Brasileiro de Pesquisas F\'\i sicas\\
22290-180, Rio de Janeiro-RJ, Brazil\\
 {}$^{(b)}$ Departamento de F\'\i sica, Universidade Federal de
Campina Grande \\
58109-790 Campina Grande, PB, Brazil\\
 {}$^{(c)}$ Departamento de
F\'\i sica, Universidade Federal da
Para\'{\i}ba \\
58051-970, Jo\~ao Pessoa, PB, Brazil}

\begin{abstract}
We consider the supersymmetric quantum mechanics(SUSY QM) with three-component
normal modes for the Bogomol'nyi-Prasad-Sommerfield (BPS) states. An explicit
form of the SUSY QM matrix superpotential is presented and the corresponding
three-component bosonic zero-mode eigenfunction is investigated.

\vspace{1.0cm}



\vspace{4cm}
Permanent address of RLR:
Unidade Acad\^emica de Educa\c c\~ao, Universidade Federal
de Campina Grande, Cuit\'e - PB, 58.175-000, Brazil. \\

E-mails: rafael@df.ufcg.edu.br, aerlima@df.ufcg.edu.br,
emello@fisica.ufpb.br, valdir@fisica.ufpb.br

\end{abstract}

\maketitle

\newpage

\section{Introduction}

\paragraph*{}
Supersymmetric quantum mechanics(SUSY QM) has provided a possibility
to solve analytically some non-relativistic quantum systems. The
simplest model in this framework was proposed by Witten\cite{W} in
the early eights of the last Century. After this pioneered work, the
methods of SUSY QM have quickly developed and some applications
arised. Some of these applications include the problems related with
the superpotential \cite{vvr02,vachas04}, whose generalization to
the case of a matrix superpotential was done a long time ago.   The
SUSY QM has also applications in the non-relativistic quantum
context \cite{Fred}-\cite{RPV98} and in the case involving two or
three fields in (1+1)-dimensional model. Others applications include
some results connected with self-adjoint extensions \cite{Falomir}
and superpotential matrix \cite{Kumar}, among others.

The classical configurations with domain wall solutions  are
bidimensional structures in (3+1)-dimensions
\cite{Jac}-\cite{Weinberg}. They are static, non-singular,
classically stable Bogomol'nyi \cite{Bogo} and Prasad-Sommerfield
\cite{PS} (BPS)  soliton (defect) configurations, with finite
localized energy associated with a real scalar field potential
model.

The BPS states are classical configurations that satisfy first and second order
differential equations. In a context that stresses the connection with BPS-bound
states\cite{guila02,GER07}, domain walls have been exploited.

Recently, the stability and metamorphosis of BPS states have been
investigated \cite{Shif01b}, using the framework of SUSY QM, with a
detailed analysis of a 2-dimensional $N=2-$Wess-Zumino model in
terms of two chiral superfields and composite dyons in
$N=2$-supersymmetric gauge theories\cite{KS04}. Also, the
superpotential associated with the linear classical stability of the
static solutions for systems with one real scalar field in
(1+1)-dimensions were discussed in the literature
\cite{vvr02,vachas04}. However, for solitons associated with three
coupled scalar fields there are no general rules for finding
analytic solutions since the nonlinearity in the potential leads to
an increasing of the difficulties to solve the BPS and field
equations.

This paper is organized as follows: In Section II, we discuss SUSY algebra with
topological charge. In Section III, we consider a SUSY model for two coupled scalar
fields. In Section IV, we present the
BPS configurations for three coupled scalar fields. In Section V,
we define the BPS mass bound of the energy and discuss the stability of BPS
states. The Schr\"odinger-like equation and also the Hessian matrix are obtained.
In Section VI, a matrix superpotential with three-component wave functions is obtained.
In Section VII, a specific potential model is investigated.
Our conclusions are presented in Section VIII.

\section{SUPERSYMMETRY ALGEBRA WITH TOPOLOGICAL CHARGE}

\paragraph*{}

Let us start with a discussion concerning central charges due to the fact that these quantities characterize SUSY. To do this, consider the potential model with one single
real scalar field $\phi,$ whose Lagragian is given by

\begin{equation}
\label{E1}
 A= \int d^2z
\frac{1}{2}\left\{\left(\partial_\mu\phi\right)^2+ \left[\bar\psi
\partial_\mu \gamma^\mu \psi-U^2(\phi)-U^{\prime}(\phi)\bar\psi
\psi\right]\right\}
\end{equation}
where $U(\phi)$ a well defined continuous function and the Majorana spinor, $\psi(z)$,
is given by

\begin{equation}
\label{E2}
 \psi(z)=\left(
\begin{array}{cc}
\psi_+(z) \\
\psi_-(z)
\end{array}\right).
\end{equation}
 
In this case, the conserved SUSY current can be written as

\begin{equation}
\label{E3}
S^\mu_\beta= \left(\partial_\alpha\phi\right)
\left(\gamma^\alpha\gamma^\mu\right)_{\beta\xi} \psi_\xi +
U(\phi)\gamma^\mu_{\beta\xi}\psi_\xi.
\end{equation}

Therefore, the topological SUSY charge is given by

\begin{equation}
\label{E4} Q_\beta= \int S^0_\beta dz,
\end{equation}
and, then, we can write

\begin{equation}
\label{E5} Q_+= \int dz\left[\left(\partial_0\phi +
\partial_1\phi\right)\psi_+ - U(\phi)\psi_-\right],
\end{equation}

\begin{equation}
\label{E6} Q_-= \int dz\left[\left(\partial_0\phi -
\partial_1\phi\right)\psi_- + U(\phi)\psi_+\right].
\end{equation}

In (1+1)-dimensions the SUSY algebra becomes

\begin{equation}
\label{E6a} Q^2_+= P_+= P_0 + P_1, \quad Q^2_-= P_-= P_0 - P_1
\end{equation}
and

\begin{equation}
\label{E6b} Q_+Q_- + Q_-Q_+=0
\end{equation}
where
$$ [\psi_-(y),\psi_-(x)]_+=\delta(y-x),
\quad [\psi_+(y),\psi_+(x)]_+=\delta(y-x),
$$

$$
[\psi_-(x),\psi_+(x)]_+=0.
$$

In a field theory without soliton solutions these equations are
satisfied. However, in a field theory with soliton solutions these
equations are not satisfied because the surface terms for a soliton
solution are different from zero, and as a consequence

\begin{equation}
\label{E7a} Q_+Q_- + Q_-Q_+=  \int^{+\infty}_{-\infty}
dz\frac{\partial}{\partial x} (2\Gamma(\phi)),
\end{equation}
with the superpotential satisfying the relation

\begin{equation}
\label{E8} \Gamma^\prime(\phi)= \frac{d}{d\phi} \Gamma= U(\phi).
\end{equation}
Note that the right hand side of Eq. (\ref{E7a}) is a scalar, which
corresponds exactly to the central charge. Thus,
the Bogomol'nyi classical
bound, for a single particle with mass $m_0$, at rest, which means that, 
$P_+=P_-=m_0$, becomes

\begin{equation}
\label{E11} m_0\geq\left|\int^{+\infty}_{-\infty} dz
\frac{\partial}{\partial z}\Gamma(\phi)\right| =\left|\Gamma[M_j]-
\Gamma[M_i]\right|,
\end{equation}
where $M_i$ and $M_j$ represent the vacuum states. It is worth calling attention 
to the fact that this inequality remains valid for soliton and antisoliton
solutions at one-loop order.

\section{SUSY FROM TWO COUPLED SCALAR FIELDS}

\paragraph*{}

Let us write the potential $V(\phi_j)$ in the following SUSY
form, analogous to the case with one single field only,
\begin{equation}
\label{VSS} V(\phi_j)=\frac
12\left(U_1^2(\phi_j)+U_1^2(\phi_j)\right), \quad
U_i(\phi_j)=U_i(\phi_1,\phi_2).
\end{equation}
Thus, the $N=1$ algebra can be discussed by investigating the SUSY Lagrangian 
density in (1+1)-dimensions with the following form

\begin{eqnarray}
{\cal L}&=& \frac{1}{2}\left(\partial_\mu\phi_1\right)^2 +
\frac{1}{2}\left(\partial_\mu\phi_2\right)^2 + \frac{1}{2}\bar\psi_1
\left(i\gamma^\mu\partial_\mu-
\frac{\partial U_1}{\partial \phi_1}\right)\psi_1 \nonumber\\
&& -\frac{1}{2}U_1^2(\phi_j) -\frac{1}{2}U_2^2(\phi_j) +
\frac{1}{2}\bar\psi_2\left(i\gamma^\mu\partial_\mu -
\frac{\partial U_2}{\partial \phi_2}\right)\psi_2 \nonumber\\
&& -\frac{1}{2} \frac{\partial U_1}{\partial\phi_2} \bar\psi_1
\psi_2-\frac{1}{2} \frac{\partial U_2}{\partial\phi_1} \bar\psi_2\psi_1
\end{eqnarray}
where $\psi_1$ and $\psi_2$ are Majorama spinors. In this framework, the SUSY 
current is given by

\begin{equation}
S^\mu= (\partial_\alpha\phi_1)\gamma^\alpha\gamma^\mu\psi_1 + iU_1(\phi_j) \gamma_\mu\psi_1 + (\partial_\beta\phi_2)\gamma^\beta\gamma^\mu\psi_2
+ iU_2(\phi_j)\gamma^\mu\psi_2,
\end{equation}
and therefore, the conserved supercharges can be expressed as

\begin{eqnarray}
\label{E4} Q_\pm&=& \frac{1}{\sqrt 2}\int dz\left\{(\partial_0\phi_1
\pm \partial_1\phi_1)
\psi_\pm \mp U_1(\phi_J)\psi_\mp\right\} \nonumber\\
&&+ \frac{1}{\sqrt 2}\int dz\left\{(\partial_0\phi_2 \pm
\partial_1\phi_2)\psi_{\pm} \mp U_2(\phi_j)\psi_{\mp}\right\}.
\end{eqnarray}

On the other hand, the superpotential $W(\phi_j)$ satisfy

\begin{equation}
\label{CSP2}
\frac{\partial W}{\partial\phi_1}=U_1(\phi_j), \quad
\frac{\partial W}{\partial\phi_2}=U_2(\phi_j)
\end{equation}
which leads to the value for a Bogomol'nyi minimum energy.

\section{Configuratios with three coupled scalar fields}

\paragraph*{}

In this section, we consider classical soliton solutions with
three coupled real scalar fields, $\phi_ {j}, (j=1,2,3)$, in (1+1)-dimensions included in bosonic sector and explain the equality of topological and central charges, $\psi_i=0.$ The soliton solutions are static, nonsingular, classically stable and finite localized energy
solutions of the field equations. The Lagrangian density for such nonlinear system in
the natural system of units $(c=\hbar=1)$, in a (1+1)-dimensional space-time, with Lorentz invariance, is written as

\begin{equation}
{\cal L}\left(\phi_j, \partial_{\mu}\phi_j\right)
= \frac{1}{2}\sum^3_{j=1}\left(\partial_{\mu}\phi_j\right)^2 -V(\phi_j),
\end{equation}
where $\partial_{\mu}=\frac{\partial}{\partial z^{\mu}}, \quad
z^{\mu}=(t,z) $ with $ \mu =0, 1, \quad \phi_j =\phi_j(t,z)$ and $\eta^{\mu\nu}=diag(+,-)$ is the metric tensor. Here, the potential
$V(\phi_j)=V(\phi_1,\phi_2,\phi_3)$ is a positive definite function of
$\phi_j$. The general classical configurations obey the following equation

\begin{equation}
 \frac{\partial^2}{\partial t^2}\phi_j - \frac{\partial^2}{\partial z^2}\phi_j
+\frac{\partial }{\partial \phi_j}V=0,
\end{equation}
which, for static soliton solutions, is equivalent to the following
system of nonlinear second order differential equations

\begin{equation}
\label{EMG}
 \phi_j^{\prime\prime}=\frac{\partial }{\partial \phi_j}V, \quad (j=1,2,3),
\end{equation}
where prime denotes differentiation with respect to the space
variable.

There is in literature, a trial orbit method for finding
static solutions of Eq.(\ref{EMG}), for certain positive potentials. This constitutes
what is termed the "trial and error" technique \cite{Raja}. This method has many limitations, notably the need to choose trial orbits. Solutions had to be obtained by
ingenuity combination with trial and error rather than by systematic derivation
from the field equations. In this paper we will use the trial orbit method for the first order differential equations associated with three real scalar fields, differently from what was done by Rajaraman\cite{Raja}, who applied this method to the equation of motion.

Let us assume that the trial orbit is given by

\begin{equation}
\label{TO}
 G(\phi_1, \phi_2, \phi_3)=0.
\end{equation}

Thus, we have

\begin{equation}
\label{TO-2}
\frac{d}{dz}G(\phi_1, \phi_2, \phi_3)=\sum_{i=1}^3\frac{\partial
 G}{\partial\phi_i}\phi_i^{\prime}=0.
\end{equation}

Taking Eqs. (\ref{TO}) and (\ref{TO-2}) into account, we can get the constant
coefficients in such trial orbit by substitution of the vacuum and the BPS states into these equations.

\section{Linear Stability}

\paragraph*{}

Since the potential $V(\phi_j)$ is positive, it can be written in the
square form analogous to the case in which we have just one single
field\cite{vvr02}, as

\begin{equation}
\label{EP}
  V(\phi_j)=
V(\phi_1,\phi_2,\phi_3)=\frac{1}{2}\sum^3_{j=1}U_j^2(\phi_1,\phi_2,\phi_3),
\quad U_j(\phi_1,\phi_2,\phi_3) \equiv\frac{\partial
W}{\partial\phi_j},
\end{equation}
where $W$ is the superpotential associated with the three fields.

Therefore, we can write the total energy given by

\begin{equation}
\label{ET3}
E=\int_{-\infty}^{+\infty} dz\frac 12\left[\left(\phi_1^{\prime}\right)^2
+\left(\phi_2^{\prime}\right)^2 +\left(\phi_3^{\prime}\right)^2 +
2V(\phi,\chi)\right],
\end{equation}
in the  BPS form, which consists of a sum of squares and surface terms, as

\begin{equation}
\label{ETG} E=\int_{-\infty}^{+\infty} dz \left(\frac
12(\phi_1^{\prime}-U_1)^2+\frac 12(\phi_2^{\prime}-U_2)^2+ \frac 12
(\phi_3^{\prime}-U_3)^2+\frac{\partial}{\partial z}W \right).
\end{equation}

Note that the first three terms are always positive and thus,
the lower bound of the energy is given by the fourth term, which means that

\begin{equation}
\label{Ebogo}
E\geq\left|\int_{-\infty}^{+\infty} dz
\frac{\partial}{\partial z}W[\phi_1(z), \phi_2(z), \phi_3(z)]\right|,
\end{equation}
where the superpotential $W=W[\phi_1(z), \phi_2(z), \phi_3(z)]$ will be discussed in
what follows. The BPS mass bound of the energy which results in a topological charge
is given by

\begin{equation}
\label{EBPS}
E_{BPS}=T_{ij}=|W[M_j]-W[M_i]|,
\end{equation}
where $M_i$ and $M_j$ represent the BPS vacuum states and are the extrema of $W.$
In this case the BPS states satisfy the following set of first order differential equations

\begin{equation}
\label{CBPS}
  \phi_j^{\prime}=U_j(\phi_1,\phi_2,\phi_3).
\end{equation}

Now, let us analyze the classical stability of the soliton solutions
in this nonlinear system, taking into consideration
small perturbations around $\phi_j (z) (j=1,2,3)$, namely, $\eta_{j}$. Thus,
we can write the classical solution of the system as

\begin{equation}
  \phi_j(t,z)=\phi_j(z)+\eta_j(t,z), \quad (j=1,2,3).
\end{equation}

We can expand the fluctuations $\eta_j(t,z)$ in terms of
the normal modes, in the following way

\begin{equation}
\eta_j (t,z) = \sum_n \epsilon_{j,n} \eta_{j,n} (z) e^{i\omega_{j,n} t},
\quad \omega_{1,n}=\omega_{2,n}=\omega_{3,n}=\omega_n,
\end{equation}
where $\epsilon_{j,n}$ are real constant coefficients.
Thus, the stability equation for the fields turns into a Schr\"odinger-like equation
for a three-component eigenfunction $\Psi_{n}$,

\begin{equation}
\label{SE}
{\cal H} \Psi_{n}
= \omega_{n}^2\Psi_{n}, \quad n=0, 1, 2, \cdots,
\end{equation}
where

\begin{equation}
\label{EFH1}
{\cal H}=\left(
\begin{array}{ccc}
-\frac{d^2}{dz^2} +\frac{\partial^2}{\partial\phi_1^2}V &
\frac{\partial^2}{\partial \phi_1\partial\phi_2}V
 &
\frac{\partial^2}{\partial \phi_1\partial\phi_3}V \\
 \frac{\partial^2}{\partial \phi_2\partial\phi_1}V &
-\frac{d^2}{dz^2} +\frac{\partial^2}{\partial\phi_2^2}V &
+\frac{\partial^2}{\partial \phi_2\partial\phi_3}V \\
\frac{\partial^2}{\partial \phi_3\partial\phi_1}V &
\frac{\partial^2}{\partial \phi_3\partial\phi_2}V &
-\frac{d^2}{dz^2} +\frac{\partial^2}{\partial\phi_3^2}V
\end{array}\right)_{\vert\phi_j =\phi_j(z)}
\equiv-{\bf I}\frac{d^2}{dz^2}+V_F(z),
\end{equation}
with {\bf I} being the (3x3)-dentity matrix and $V_F(z)$ the (3x3) fluctuation Hessian matrix. The excited modes are, thus, given by

\begin{equation}
\label{EFH} \Psi_{n}(z)=\left(
\begin{array}{ccc}
\eta_{1,n}(z) \\
\eta_{2,n}(z) \\
\eta_{3,n}(z)
\end{array}\right).
\end{equation}
Since $V_F(z)$ is a symmetric matrix and ${\cal H}$ is Hermitian, thus the
eigenvalues $\omega^2_n$ of ${\cal H}$ are real.

The Schr\"odinger-like equation (\ref{SE}) and the Hessian matrix
$V_F(z)$ in Eq. (\ref{EFH1}) are obtained by taking a Taylor
expansion of the potential $V(\phi_j)$ in terms of $\eta_j(t,z)$ and
retaining the first order terms in the equations of motion.

\section{Potential model with three scalar fields}

\paragraph*{}

As an application of this formalism, let us consider the following
potential

\begin{eqnarray}
\label{EP3} V=&&V(\phi_1,\phi_2,\phi_3)= \frac
12\left(\lambda\phi_1^{2}+\alpha\phi_2^2
+\alpha\phi_3^2 -\frac{m^2}{\lambda}\right)^2\nonumber\\
+&&\frac
12\left(-\alpha\phi_1\phi_2+\beta_2\phi_3^2-\beta_2\right)^2
\nonumber\\
+&&\frac 12\phi_3^2
\left(-\alpha\phi_1+2\beta_2\phi_2+\alpha\beta_1\right)^2,
\end{eqnarray}
where $\alpha>0$ and $\beta_i\geq 0.$ This is a generalized potential for three scalar fields which
was constructed from the potential discussed recently \cite{GER07}, for two scalar fields. Note that the
symmetry $Z_2$x$Z_2$ is preserved only if $\phi_2=0$ or if $\beta_1=\beta_2=0$.

The corresponding superpotential in a field theory model is given by

\begin{equation}
\label{SM3} W(\phi_j)=\frac{m^2}{\lambda}\phi_1-
\frac{\lambda}{3}\phi_1^3-\alpha\phi_1\phi_2^2
-\frac{\alpha}{2}\phi_1\phi_3^2+\beta_2\phi_2\phi_3^2-\beta_2\phi_2+
\frac 12\alpha\beta_1\phi_3^2.
\end{equation}

It is required that $\phi_j$, satisfy the BPS state conditions, which are expressed
by the following equations

\begin{eqnarray}
\label{EBPS3}
  \phi_1^{\prime}&&=-\lambda\phi_1^{2}-\alpha\phi_2^2+\frac{m^2}{\lambda}
-\alpha\phi_3^2  \nonumber\\
\phi_2^{\prime}&&=-2\alpha\phi_1\phi_2+\beta_2\phi_3^2-\beta_2 \nonumber\\
\phi_3^{\prime}&&=\phi_3(-\alpha\phi_1+2\beta_2\phi_2+\alpha\beta_1)
\end{eqnarray}
and the superpotential $W(\phi_j)$ satisfy
$\frac{\partial W}{\partial\phi_j}=U_j (j=1,2,3).$

Note that the BPS states saturate the lower bound, so that
$E_{BPS}=|W_{ij}|$ is the central charge of the realization of
$N=1$ SUSY in (1+1)-dimensions. Thus, the vacua are determined by the extrema of the
superpotential. Therefore, the condition

\begin{equation}
\frac{\partial W}{\partial\phi_j}=0, \quad j=1, 2, 3
\end{equation}
provides the vacuum states $M_i=(\phi_{1\hbox{v}}, \phi_{2\hbox{v}},
\phi_{3\hbox{v}})$ whose values must satisfy the following equations

\begin{eqnarray}
\label{CV} {}&&-\lambda\phi_1^{2}-\alpha\phi_2^2+\frac{m^2}{\lambda}
-\frac 12 \alpha\phi_3^2=0  \nonumber\\
{}&&-2\alpha\phi_1\phi_2+\beta_2\phi_3^2-\beta_2=0 \nonumber\\
{}&&-\alpha\phi_1+2\beta_2\phi_2+\alpha\beta_1=0.
\end{eqnarray}

In order to obtain an explicit form of the vacuum states,
let us consider the cases $\phi_{2\hbox{v}}=\phi_{3\hbox{v}}=0$ and
$\phi_{1\hbox{v}}=\phi_{3\hbox{v}}=0,$ respectively. Thus,
we obtain the four vacuum states, which are given by

\begin{eqnarray}
\label{V1234}
  M_1&&=\left(-\frac{m}{\lambda},0,0\right) \nonumber\\
  M_2&&=\left(\frac{m}{\lambda},0,0\right)\nonumber\\
  M_3&&=\left(0,-m{\sqrt{\frac{1}{\lambda\alpha}}},0\right) \nonumber\\
  M_4&&=\left(0, m{\sqrt{\frac{1}{\lambda\alpha}}},0\right).
\end{eqnarray}

It is easy to verify that these vacuum states are satisfied by the equations
given in (\ref{CV}), for $\beta_2=0$ and $\alpha\lambda>0.$ Now, let consider
$\phi_{1\hbox{v}}=\beta_1$ and $\phi_{2\hbox{v}}=0$ in Eq.(\ref{CV}). In this case,
we obtain two additional vacuum states, which are

\begin{eqnarray}
\label{V56}
  M_5&&=\left(\beta_1, 0, \sqrt{\frac{2}{\alpha}\left(\frac{m^2}{\lambda}-
\lambda\beta_1^2\right)}\right) \nonumber\\
  M_6&&=\left(\beta_1, 0, -\sqrt{\frac{2}{\alpha}\left(\frac{m^2}{\lambda}-
\lambda\beta_1^2\right)}\right),
\end{eqnarray}
for $\beta_2=0, -\frac{m}{\lambda}<\beta_1<\frac{m}{\lambda}$ and
$\alpha\neq 0.$ Therefore, in this case the components of the tension are

\begin{eqnarray}
\label{ETT}
T_{12}&&=T_{21}=\frac 43\frac{m^3}{\lambda^2}\nonumber\\
T_{13}&&=T_{31}= T_{24}=T_{42}=T_{41}=T_{14}=T_{23}=T_{32}
=\frac 23\frac{m^3}{\lambda^2} \nonumber\\
T_{15}&&=T_{51}= T_{25}=T_{52}=T_{16}=T_{61}= T_{26}=T_{62}=|\frac
23\frac{m^3}{\lambda^2}-
\beta_1(\frac{m^2}{\lambda}-\frac{\lambda}{2}\beta_1^2)| \nonumber\\
T_{34}&&=T_{43}= 0=T_{56}=T_{65}\nonumber\\
T_{35}&&=T_{53}= T_{45}=T_{54}=T_{36}=T_{63}=|
\frac{\lambda}{3}\beta_1^3-\frac{m^2}{\lambda}\beta_1|.
\end{eqnarray}

From the results given by Eq.(\ref{ETT}), we see that the potential
presents two non-topological
sectors, which are non-BPS sectors, namely, $T_{34}$ and $T_{56},$
and twelve BPS topological sectors.

Now, let us specialize to the (3x3)-matrix superpotential, {\bf W},
with $\beta_2=0,$ which is given by

\begin{equation}
\label{spg} \hbox{{\bf W}}=\left(
\begin{array}{ccc}
 2\lambda\phi_1 &  \alpha\phi_2 &  \alpha\phi_3\\
 \alpha\phi_2 & \alpha\phi_1 & 0 \\
\alpha\phi_3 & 0 & \alpha\beta_1-\alpha\phi_1
\end{array}\right)_{\vert\phi=\phi(z),\chi=\chi(z)}.
\end{equation}
This superpotential satisfies the Ricatti equation associated with the
non-diagonal fluctuation Hessian matrix, $V_{F}(z)$, which is written as

\begin{equation}
\label{ER} \hbox{{\bf W}}^2+\hbox{{\bf W}}^{\prime}=V_{F}(z)=\left(
\begin{array}{ccc}
 V_{F11}(z) &  V_{F12}(z)&  V_{F13}(z)\\
 V_{F12}(z) & V_{F22}(z) &  V_{F23}(z)\\
V_{F13}(z) & V_{F23}(z) &  V_{F33}(z)
\end{array}\right)_{\vert\phi=\phi(z),\chi=\chi(z)},
\end{equation}
where the elements of $V_{F}(z)$, denoted by $V_{Fij}(z)$,are given by the following relations

\begin{eqnarray}
\label{ET}
V_{F11}&&=6\lambda^2\phi_1^2+
\alpha^2(4\phi_2^2+\phi_3^2)+2\lambda\left(\alpha\phi_2^2+
\frac 12 \alpha\phi_3^2-\frac{m^2}{\lambda}\right)\nonumber\\
 V_{F22}&&=6\alpha^2\phi_2^2+ 2\alpha(2\alpha+\lambda)\phi_1^2+
(\alpha^2+4\beta_2^2)\phi_3^2
-2\alpha\frac{m^2}{\lambda}\nonumber\\
V_{F12}&&=V_{21}=4\lambda\alpha\phi_1\phi_2+
8\alpha^2\phi_1\phi_2
-2\alpha(2\beta_2\phi_3^2-\beta_2)\nonumber\\
V_{F13}&&=V_{31}=2\alpha(\lambda+\alpha)\phi_1\phi_3-
4\alpha\beta_2\phi_2\phi_3 -2\alpha^2\beta_1\phi_3\nonumber\\
V_{F33}&&=\frac 32 \alpha^2\phi_3^2+ 6\beta^2_2 \phi_3^2-
2\beta_2(2\alpha\phi_1\phi_2+\beta_2)+
(-\alpha\phi_1+2\beta_2\phi_2+\alpha\beta_1)^2
\nonumber\\
V_{F32}&&=V_{F23}=2\alpha^2\phi_2\phi_3
+4\phi_3\beta_2(-\alpha\phi_1+2\beta_2\phi_2^2+\alpha\beta_1).
\end{eqnarray}
The Ricatti equation (\ref{ER}) only holds for the BPS states. According
to Witten's model\cite{W,Fred}, we have

\begin{equation}
\label{A+-} {\cal A}^{\pm}=\pm\hbox{{\bf I}} \frac{d}{dz}+\hbox{{\bf
W}}(z), \quad \Psi_{\hbox{SUSY}}^{(n)}(z)=\left(
\begin{array}{cc}
\Psi_-^{(n)}(z) \\
\Psi_+^{(n)}(z)
\end{array}\right)_{1\hbox{x}6},
\end{equation}
where  $\Psi_{\pm}^{(n)}(z)$ are three-component eigenfunctions. In
this case, the graded Lie algebra of the SUSY QM for the BPS
states may be realized as

\begin{equation}
\label{E40} H_{SUSY} = [Q_- ,Q_+ ]_+ = \left(
\begin{array}{cc}
{\cal A}^+ {\cal A}^- & 0 \\
0 & {\cal A}^- {\cal A}^+
\end{array}\right)_{6\hbox{x}6}= \left(
\begin{array}{cc}
{\cal H}_- & 0 \\
0 & {\cal H}_+
\end{array}\right),
\end{equation}

\begin{equation}
\label{41} \left[H_{SUSY} , Q_{\pm}\right]_- = 0 = (Q_-)^2 =
(Q_+)^2,
\end{equation}
where $Q_{\pm}$ are the (6x6) supercharges of the Witten model and is given by

\begin{equation}
\label{SC} Q_- = \sigma_-\otimes{\cal A}^-, \quad Q_+=Q_-^{\dagger}=
\left(
\begin{array}{cc}
0 & {\cal A}^+ \\
0 & 0
\end{array}\right)=\sigma_+\otimes{\cal A}^+,
\end{equation}
with the intertwining operators, ${\cal A}^{\pm}$, given in terms of
(3x3)-matrix superpotential, Eq.(\ref{A+-}), and
$\sigma_{\pm}=\frac 12(\sigma_1\pm i\sigma_2),$ with $\sigma_1$ and
$\sigma_2$ being Pauli matrices. Note that the bosonic sector of
$H_{SUSY}$ is exactly the fluctuating operator given by ${\cal
H}_-={\cal H}=-{\bf I}\frac{d^2}{dz^2} +{\bf V}_{F}(z),$ where ${\bf
V}_-={\bf V}_{F}(z)$ is the non-diagonal fluctuation Hessian matrix. The
supersymmetric fluctuation partner operator of ${\cal H}_-$ is

\begin{equation}
\label{SP} {\cal H}_+={\cal A}^- {\cal A}^+={\cal A}^+ {\cal
A}^-+[{\cal A}^-,{\cal A}^+] ={\cal H}_--{\bf W}^{\prime}(z),
\end{equation}
so that the SUSY partner is given by ${\bf V}_{+}={\bf V}_- -{\bf
W}^{\prime}(z).$

The  Ricatti equation given by
(\ref{ER}) is reduced to a set of first-order coupled differential
equations. In this case, the superpotential is not necessarily defined as
$W(z)=\frac{1}{\psi_-^{(0)}}\frac{d}{dz}\psi_-^{(0)}(z),$ as in the case of
a system described by a one-component wave function in the framework of SUSY QM\cite{W,Fred}.

Therefore,  as the zero-mode is associated with a three-component
eigenfunction, $\Psi_-^{(0)}(z)$,  one may write the matrix
superpotential in the form\cite{rafa01}

\begin{equation}
\frac{d}{dz}\Psi_-^{(0)}(z)= {\bf
W}\Psi_-^{(0)}(z),
\end{equation}
from which we find the following zero mode eigenfunction

\begin{equation}
\label{AF}
  \Psi_{-}^{(0)}=\left(
\begin{array}{ccc}
U_1(\phi_i) \\
 U_2(\phi_i) \\
U_3(\phi_i)
\end{array}\right),
\end{equation}
where $U_i \quad (i=1, 2, 3)$ are given by the BPS states (\ref{EBPS3}).

Now, let us show that the $\omega_n{^2}$' s are non-negative.
To do this, consider the bilinear form
of ${\cal H}$ given by

\begin{equation}
{\cal H}=
 {\cal A}^+ {\cal A}^-,
\end{equation}
where

\begin{equation}
{\cal A}^{-}=({\cal A}^+)^{\dag}=\left(
\begin{array}{ccc}
a_1^- & {\cal A}^-_{12}
& {\cal A}^-_{13} \\
{\cal A}^-_{21} & a^-_2
& {\cal A}^-_{23} \\
{\cal A}^-_{31} & {\cal A}^-_{32} & a_3^-
\end{array}\right)=\left(
\begin{array}{ccc}
a_1^- & 0 & 0 \\
0 & a^-_2 & 0 \\
0 & 0 & a_3^-
\end{array}\right)+{\cal R}(\phi_i),
\end{equation}
with the obvious identification of the elements of ${\cal R}(\phi_i)$ and the
following expressions for the operators that appear in the
analysis of classical stability associated with a single field
\cite{vvr02}

\begin{eqnarray}
 a^-_1&&=-\frac{d}{dz}+2\lambda\phi_1, \nonumber\\
a^-_2&&=-\frac{d}{dz}+\alpha\phi_1, \nonumber\\
a^-_3&&=-\frac{d}{dz}+\alpha\phi_1-\alpha\beta_1,
\end{eqnarray}
where
$$ {\cal A}^-_{12}=\alpha\phi_2={\cal A}^-_{21}, \quad {\cal
A}^-_{13}=\alpha\phi_3={\cal A}^-_{31},\quad
 {\cal A}^-_{23}= 0={\cal
A}^-_{32}.$$

Since $a^+_j=(a^-_j)^{\dagger}$ and hence ${\cal A}^+=({\cal A}^-)^{\dag},$ we
find

\begin{equation}
 ({\cal A}^+ {\cal A}^-)_{jj}
=-\frac{d^2}{dz^2}+\frac{\partial^2}{\partial\phi^2_j}V,
\end{equation}
which are exactly the diagonal elements of ${\cal H}$. It is worth calling attention to the fact that the linear stability is satisfied, which means that

\begin{equation}
 \omega^2_n =
<{\cal H}>=< {\cal A}^+ {\cal A}^->=
({\cal A}^-{\tilde\Psi}_{n})^{\dag}({\cal A}^-{\tilde\Psi}_{n})\geq 0,
\end{equation}
and therefore the $\omega_n{^2}$' s are non-negative.

\section{Projections on scalar fields}

\paragraph*{}

Let us now consider a projection on the $(\phi_1, \phi_2)$ plane in
order to find an explicit form of domain walls using the trial orbit
method. In this case, if we choose $\phi_3=0$ in Eq. (\ref{EBPS3})
and the following trial orbit

\begin{equation}
G(\phi_1, \phi_2)=c_1\phi_1^2+c_2\phi_2^2+c_3=0,
\end{equation}
we get from $\frac{dG}{dz}=\frac{\partial
G}{\partial\phi_1}\phi_1^{\prime}+\frac{\partial
G}{\partial\phi_2}\phi_2^{\prime}=0$ and using the BPS states
(\ref{EBPS3}), that $c_1=1, c_2=\frac{\alpha}{2(\lambda -
\alpha)}$ and $c_3=-\frac{m^2}{\lambda^2}.$ Thus, the resulting
elliptical orbit is

\begin{equation}
\phi_1^2+\frac{\alpha}{(\lambda -
2\alpha)}\phi_2^2=\frac{m^2}{\lambda^2}
\end{equation}
or

\begin{equation}
\frac{\lambda^2}{m^2}\phi_1^2+\frac{\lambda^2}{2m^2}\phi_2^2=1,
\end{equation}
for $\alpha=\frac{\lambda}{4}.$ These provide the following BPS
solutions

\begin{eqnarray}
\label{pbps}
\phi_1(z)&&=\frac{m}{\lambda}\tanh(\frac m2 z)\nonumber\\
\phi_2(z)&&=\pm\sqrt{2}\frac{m}{\lambda}\mbox{sech}(\frac m2 z)\nonumber\\
\phi_3=0,
\end{eqnarray}
which connect the vacua $(\frac{m}{\lambda}, 0,0)$ and $(-\frac{m}{\lambda}, 0,0).$ Note that

$$
\phi_1\rightarrow \pm\frac{m}{\lambda}, \quad \hbox{for}\quad
z\rightarrow\pm\infty
$$

$$
\phi_2\rightarrow 0, \quad \hbox{for} \quad z\rightarrow\pm\infty.
$$

This result corresponds to the same one obtained recently\cite{GER07} for BPS solutions when two scalar fields are taken into account. 

In this situation, the matrix superpotential in SUSY QM, ${\bf W}(z)$, becomes

\begin{equation}
\label{E34a} {\bf W}(z)=-\frac{m}{2}\left(
\begin{array}{ccc}
4\tanh(\frac m2 z)  & \pm\sqrt{2}sech(\frac m2 z)&0 \\
 \pm\sqrt{2}\mbox{sech}(\frac m2 z) &
\tanh(\frac m2 z) &0\\
0&0&4
\end{array}\right),
\end{equation}
where we have used the BPS states in terms of a projection on the
$(\phi, \chi)$ plane. This provides the following bosonic zero-mode

\begin{equation}
\label{ZM2}
\frac{d}{dz}\Psi_-^{(0)}(z)= {\bf
W}\Psi_-^{(0)}(z)\Rightarrow\Psi_{-}^{(0)}(z)=N\left(
\begin{array}{ccc}
sech^2(\frac m2 z) \\
\pm\sqrt{2}\tanh(\frac m2 z) \mbox{sech}(\frac m2 z) \\
0
\end{array}\right),
\end{equation}
for $\alpha=\frac{\lambda}{4},$ where $N$ is the normalization constant.
However, in Ref. \cite{GER07} the relation between $\lambda$ and $\alpha$ is given by $\alpha=\frac{\lambda}{2}.$

\section{Conclusions}

\paragraph*{}

In this paper, we considered the classical stability analysis for
BPS domain walls associated with a potential model of three
coupled real scalar fields, which obeys the non-ordinary
supersymmetry (SUSY). The approach of effective quantum mechanics provides a realization of SUSY algebra in the three-domain wall
sector of the non-relativistic formalism.

The components of the tension given in (\ref{ETT}) were deduced from the charge
central properties in the model that present $N=1$ SUSY.
From a three-field specific potential model we found two null tensions which
correspond to non-topological sectors, and other topological sectors, which depend
on the manifold of vacuum states, $T_{ij}=|W[M_j]-W[M_i]|,$ where $M_i$ and $M_j$
represent the vacuum states.

We have shown that the positive potentials with a square form lead
to three-component non-negative normal modes $\omega_n{^2} \geq
0$, analogous to the case with a single field \cite{vvr02}, so
that the linear stability of the Schr\"odinger-like equations is
ensured.

We have seen that domain walls associated with the three-field
potentials have features that are not present in the one-field models. The
BPS states which connect the vacua exist within the stability domain
and minimize the energy.  Thus, they provide a realization of the
supersymmetric quantum mechanical algebra for three-component
eigenfunctions. From the stability equation, we have found an
expression for the matrix superpotential, satisfying the Ricatti
equation, within the context of unidimensional quantum mechanics.

We also deduced an (3x3)-matrix explicit form of the SUSY QM
superpotential from a field-theoretic superpotential model in (1+1)-dimensions. A general three-component zero-mode eigenfunction is
deduced, but its  explicit form is found only for the projection on
the $(\phi_1, \phi_2)$ plane, $\phi_3=0,$ and for
$\alpha=\frac{\lambda}{4},$ under which the original superpotential
becomes harmonic.

Recently, in the context of a three-field potential model was considered an
hexagonal network of static classical configuration inside a
topological soliton. Also, the 1/4 BPS equations of domain wall
junction were first obtained by Gibbons and Townsend
\cite{gibbons99} and independently by Carrol {\it et al.} \cite{jun}.
We point out that the superpotential model investigated here can be
applied to implement new string junctions by extended BPS domain walls\cite{string}.

\centerline{\bf Acknowledgments}

RLR would like to  acknowledge S. Alves for hospitality at CCP-CBPF
of Rio de Janeiro-RJ, Brazil, where the part of this work was
carried out and to J. A. Hela\"yel-Neto and I. V. Vancea for many
stimulating discussions. This work was partially supported by
Conselho Nacional de Desenvolvimento Cient\'\i fico e
Tecnol\'ogico(CNPq), by Funda\c c\~ao de Apoio \`a Pesquisa do
Estado da Para\'\i ba (FAPESQ)/PRONEX/CNPq and by Funda\c c\~ao de Apoio \`a 
Ci\^encia e Tecnologia do Esp\'\i rito Santo(FAPES)/PRONEX/CNPq.

\end{document}